\documentstyle[12pt]{article}
\def\beqa{\begin{eqnarray}}
\def\eeqa{\end{eqnarray}}
\def\beq{\begin{equation}}
\def\eeq{\end{equation}}
\begin{document}
\renewcommand{\theequation}{\thesection.\arabic{equation}}
\begin{titlepage}
        \title{Double Scalar--Tensor Gravity Cosmologies}
\author{S. Capozziello\thanks{E-mail: capozziello@physics.unisa.it}~~~ and
G. Lambiase\thanks{E-mail: lambiase@physics.unisa.it} \\
 {\em Dipartimento di Scienze Fisiche "E.R. Caianiello"} \\
 {\em Universit\'a di Salerno, 84081 Baronissi (Sa), Italy.} \\
 {\em Istituto Nazionale di Fisica Nucleare, Sez. di Napoli, Italy.} \\ }
\date{\today}
\maketitle
\begin{abstract}

We investigate homogeneous and isotropic cosmological models in scalar--tensor theories
of gravity where two scalar fields are nonminimally coupled to the geometry.
Exact
solutions are found, by Noether symmetries,
depending on the form of couplings and self--interaction potentials.
An interesting feature is that we deal with the Brans--Dicke field and the
inflaton on the same ground since both are nonminimally coupled and not
distinguished {\it a priori} as in earlier models. This fact allows to
improve dynamics to get successful extended inflationary scenarios.
Double inflationary solutions are also discussed.
\end{abstract}

\thispagestyle{empty}

\vspace{20.mm}

PACS number(s): 04.50.+h, 98.80.Cq\\

\vspace{5.mm}

Keyword(s): Cosmology, Alternative Theories of Gravity, Nonminimal Coupling.

\vfill

\end{titlepage}

\section{\normalsize\bf Introduction}
\setcounter{equation}{0}

Extended gravity theories have recently assumed a
prominent role in theoretical physics
investigations since any unification scheme,
as supergravity or superstrings, in the
weak energy limit, or viable early universe cosmological models,
as extended inflation,
seem to base on them.

Besides, any effective theory, where quantum fields are taken into account in a curved
space--time, results in nonminimal couplings between geometry and matter (scalar) fields
\cite{birrel}. Furthermore, notwithstanding the fact that Einstein's general relativity
is experimentally tested with high degree of accuracy, from solar system tests to binary
pulsar observational data, it has become a peremptory necessity to consider {\it
alternative} theories of gravity. The issues is now: What kind of theory? The plethora
of them is overwhelming from higher--order gravity theories, to Kaluza--Klein
multidimensional theories, to induced--gravity theories, to gauge theories with torsion.
Several of them seem to be consistent with some quantum gravity effect in the weak
energy limit but, till now, no one can be universally considered the ''full" quantum
gravity theory.

Despite of this shortcoming, most of them have remarkable physical implications from
cosmology to particle physics. A particularly relevant role is played in inflationary
cosmology where, from the early Starobinsky model \cite{starobinsky} to the more recent
hyperextended models, non--standard theories have been widely used. In fact, the goal of
every inflationary model is to generate a brief period in which the scalar factor of the
universe, $a(t)$, increases superluminally, i.e. $a(t)>t$. If $a(t)$ grows by $e^{60}$
or more during this period, the horizon, flatness and monopole problems can be resolved.
In addition, inflation generates energy density fluctuations which may be seeds for
large scale structure formation. This features can be obtained in several alternative
theories of gravity.

However, designing a detailed microphysical model that accomplishes all of these goals
has proven to be extremely difficult and several times the extended theories have to be
adjusted to reach some partial goal.

As a general scheme, one needs a {\it grand entrance} into inflation which is a
mechanism which drives the universe into a false vacuum phase.The large, positive vacuum
density acts as an effective cosmological constant which triggers a period of a (quasi)
de Sitter expansion. Then one needs a {\it graceful exit}: a mechanism capable of
terminating the inflationary expansion, reheat the universe to a high temperature, and
restore a Friedman expansion. This second issue can result highly problematic due to the
fine--tuning of parameters which one needs to connect in the isotropy of the cosmic
microwave background (e.g. the so called ''big bubble problem" \cite{weinberg}).

Cosmological models deduced from nonminimally coupled theories of gravity (e.g.
Brans--Dicke or induced gravity) have provided schemes capable of escaping such
difficulties. Extended inflationary models could accomplish several goals which earlier
models failed and, in particular, they cure some shortcomings
of ''old" inflation.
Extended inflationary models also start when the universe is trapped in a false vacuum
state by a large energy barrier during a first--order phase transition \cite{la}. The
failure of old inflation was that the universe could never escape the false vacuum state
since the rate of tunneling through the barrier remains small compared to the
inflationary expansion rate \cite{guth}.
Extended models, in the various versions, avoid
the same failure by introducing mechanisms so that the tunneling rate exceeds the
expansion rate and, hence, the transition to the true vacuum can be
completed. For a comprehensive review, see \cite{kolb}.

The key ingredient on which extended inflation lies is the relation between the
tunneling or bubble nucleation rate, $\lambda$, and the expansion rate (the Hubble
parameter) $H$. The ratio is the dimensionless ''bubble nucleation rate" $\epsilon
=\lambda/H^4$. The false vacuum can be percolated by true vacuum bubbles only if
$\epsilon$ exceeds a critical value, $\epsilon_{crit}\simeq 0.2$. In old inflation,
$\epsilon$ is time--independent since $\lambda$  and  $H$ (which depends on the false
vacuum energy) do not vary during inflation \cite{guthwein}.

Two situations are possible: i) $\epsilon < \epsilon_{crit}$, in
which case the true vacuum bubbles never percolate and the universe
inflates forever, or ii) $\epsilon > \epsilon_{crit}$, in which
case the true vacuum percolates, but so quickly that there is
insufficient inflation to solve cosmological problems. A way to
bypass this shortcoming is to avoid bubble nucleation altogether.
{\it New inflation} and {\it chaotic inflation} utilize this approach. For
example, in new inflation, the energy barrier disappears
altogether as the universe supercools and the universe evolves
slowly but continuously from the false to true vacuum phase
\cite{linde,albrecht}. However, the model has to be fine--tuned.

Extended inflation models employ an alternative approach asking for the time variation
of $\epsilon$. Initially, $\epsilon$ is much less than $\epsilon_{crit}$ to achieve
sufficient inflation, but then it grows during inflation
to a value $\epsilon >
\epsilon_{crit}$ so that the phase transition can be completed.
Being $\lambda$ fixed
by the form of the self interacting potential, the only quantity on which
one can act is $H$. Cosmological solution with a time--varying $H$ can be easily
obtained by using modified theories of gravity like
scalar--tensor theories. A
simple extended inflationary model \cite{la} can be contructed by using the
Brans--Dicke theory \cite{brans}. The gravitational action is
\begin{equation}\label{1}
 {\cal A}=\int d^4x\sqrt{-g}\,\left[\phi R-\omega\left(\frac{\phi^{\mu}\phi_{\mu}}{\phi}
 \right)+{\cal L}_m\right]\,,
\end{equation}
 where $\phi$ is the Brans--Dicke scalar field, $\omega$ is a dimensionless parameter
and ${\cal L}_m$ is the matter Lagrangian density including all the non--gravitational
fields (from now on, we shall use natural units $\hbar=c=k_B=8\pi G=1$). The
cosmological Friedman--Robertson--Walker (FRW) equations are
\begin{equation}\label{2}
  H^2=\frac{\rho_m}{3\phi}-\frac{k}{a^2}+\frac{\omega}{6}\left(\frac{\dot{\phi}}{\phi}\right)^2
  -H\left(\frac{\dot{\phi}}{\phi}\right)\,,
\end{equation}
\begin{equation}\label{3}
  \ddot{\phi}+3H\dot{\phi}=\frac{\rho_m-3p_m}{3+2\omega}\,,
\end{equation}
 where $H=\dot{a}/a$, $\rho_m$ and $p_m$ are, respectively, the energy and pressure
densities of matter, and finally, $k=0, \pm 1$. For $p_m=-\rho_m$ and $k=0$ we have
\begin{equation}\label{4}
  \phi(t)=\left(1+\frac{\chi t}{\alpha}\right)^2\,,
\end{equation}
\begin{equation}\label{5}
  a(t)=\left(1+\frac{\chi t}{\alpha}\right)^{\omega+1/2}\,,
\end{equation}
 where $\chi$ is an integration constant connected to the energy of false vacuum,
$\alpha=(3+2\omega)(5+6\omega)/12$.

Immediately we see that for $\chi t<\alpha$, $\dot{\phi}\simeq 0$,
we have a de
Sitter solution. If, for example, $\omega >90$, one obtains the 60 $e$--foldings
necessary to solve cosmological problems of standard model. When $\chi t>\alpha$, $a(t)$
evolves as a power--law expansion and $H\sim t^{-1}$. This feature allows the successful
graceful exit since $\epsilon>\epsilon_{crit}$. Of course, the main ingredients are the
variation of Newton constant and the coupling of the geometry to the scalar field. In
other words, the model succeeds because the effective Newton constant
$G_{eff}=\phi^{-1}$ is decreasing, then $H$ is decreasing and
$\epsilon$ is increasing.

The main flaw of this model is related to the expected value of
the parameter $\omega$. In order to restore Einstein's general
relativity, we should have $\omega\to\infty$ \cite{faraoni}, then
the value of $\omega$ in constrained by the classical tests of
general relativity: light deflection and time--delay experiments
require $\omega >500$ \cite{reasenberg} while the bounds on the
anisotropy of the microwave background radiation give $\omega\leq
30$ \cite{bertschinger}. In conclusion, $\omega$ must be a
function of time in order to obtain viable models. A pure
Brans--Dicke theory is not able to yield realistic models and we
have to introduce, at least, a function $\omega =\omega(\phi)$ in
order to overcome the above difficulties. Several proposal have
been done to improve the early extended inflationary model and, in some of
them also $\lambda$ is assumed to vary \cite{kolb}. The strict condition
which implies $\lambda=$constant is a feature connected to Brans--Dicke
models as it is shown in \cite{vadas}.

In the so called {\it hyperextended inflation} \cite{accetta},
one can assume
\begin{equation}\label{6}
  \omega(\phi)=\omega_0+\omega_m\phi^m\,,
\end{equation}
 to improve the Brans--Dicke model \cite{barrow}. If $m=5$, the microwave background
bounds are satisfied \cite{liddle}. Alternatively, assuming the coupling
\begin{equation}\label{7}
  \phi=F(\varphi)\,, \quad \omega(\phi)=\frac{F(\varphi)}{2(dF/d\varphi)^2}\,,
\end{equation}
 one can get successful implementations without fine--tuning the initial conditions. In
particular, if $F(\varphi)$ is a sixth order polynomial, the big bubble problem is
avoided since the model is independent of the bubble--size distribution \cite{liddle}.

Other approaches give interesting results. For example, it is possible to include a
first- or second-order potential $V(\phi)$ for the Brans--Dicke field $\phi$ as in
induced gravity theories \cite{trester}. In this case,
the potential places constraints
on the percolation time--scale in the graceful exit and, furthermore, it can give rise to
multiple episodes of inflation which may reveal extremely useful for large scale
structure formation (e.g. super-cluster, cluster and galaxies).

Another way to escape the $\omega$--parameter constraints is to
consider a curvature--coupled inflation \cite{laycock}. Also in
this case, extended inflation results enhanced since, in some
sense, the roles of Brans--Dicke field and inflaton are mixed.
This feature allows to satisfy the solar system constraints on
$\omega$, to avoid the big--bubble problem, to construct models
with double inflationary episodes.

A more sophisticated way to bypass the graceful exit problem can be obtained by coupling
first--order phase transitions to curvature--squared inflation \cite{amendola}. The
mechanism {\it (getaway inflation)} is based on a nonminimally coupled higher--order gravity
theory where terms like $\varphi^2R^2$ appear in the usual gravity--inflaton action.
Their role is to produce an inflationary phase of the background which has a classical
end. At the same time, a stage of bubble production via semiclassical tunneling occurs
allowing useful spectra for large scale structure formation.

A final remark concerns the role which the Brans--Dicke scalar could have for dark matter in
extended inflation. Its oscillations in the various models could account for the
discrepancy between the dynamical estimate of the density of matter
in the universe,
$\Omega\simeq 0.2$, and the prediction of inflation,
$\Omega=1$ \cite{mcdonald} which seems to be confirmed
by the BOOMERANG experiment \cite{melchiorri}.

All these arguments and several more make extremely interesting to search for
cosmological solutions useful for extended inflation. A first investigation in this
sense is in \cite{barrow} where general scalar--tensor theories of gravitation were
studied in order to ''model" useful extended inflationary behaviours.
{\it Intermediate
inflationary universes} with expansion scale factor of the form
\begin{equation}\label{8}
  a(t)=a_0\exp t^p\,, \quad 0< p< 1\,,
\end{equation}
were found. These models allow to  succeed in realizing phase
transition and graceful exit.

More recently, Modak and Kamilya \cite{modak}
derived exact cosmological solutions by
the so called {\it Noether Symmetry Approach} \cite{capozziello}
in scalar--tensor gravity theories
discussing the role of the coupling function $\omega(\varphi)$ connected to the Noether
symmetry. They improved the approach in \cite{cimento}, where symmetries and solutions
were found for theories with a scalar field nonminimally coupled to gravity, by
introducing a second scalar field (the inflaton)
as in extended inflationary models.
Exponentially expanding solutions, in asymptotic region, were found and this feature
does not allow to solve the graceful exit problem also if general relativity was
asymptotically recovered.

In this paper we want to discuss, by Noether Symmetry Approach,
a further generalization
taking into account two nonminimally coupled scalar fields and their self interaction
potentials. In this way, the roles of the Brans--Dicke field and the inflaton are mixed
and both fields are taken on the same ground. This fact could be coherent with the
stochastic approach for the fundamental laws of nature since the role of the fields is
not attributed {\it a priori} \cite{ottewill}.

The paper is organized as follows. In Sect. 2, we discuss the double scalar--tensor
action, derive the equations of motion, the point-like FRW Lagrangian and the
cosmological equations. Sect. 3 is devoted to the Noether Symmetry Approach which has to
be improved for the double field case since the configuration space results enlarged.
The summary of found symmetries is given considering also the subcases where one and not
two nonminimal couplings are present. The cosmological solutions are given in Sect. 4
while the graceful exit problem is discussed in Sect. 5. Conclusions are drawn in
Sect.6.

\section{\normalsize \bf Double Scalar-Tensor Action and Equations of Motion}
\setcounter{equation}{0}

The most general action in four dimensions, where gravity is nonminimally coupled to two
scalar fields noninteracting between them, is
 \beq\label{9}
 {\cal A}=\int d^{4}x\sqrt{-g}\left[F(\varphi )R+G(\psi )R+\frac{1}{2}
 \varphi_{\mu}\varphi^{\mu}-V(\varphi
 )+\frac{1}{2}\psi_{\mu}\psi^{\mu}-W(\psi )\right]\,{,}
 \eeq
 where we have not specified the four functions $F(\varphi)$, $V(\varphi)$, $G(\psi)$,
and $W(\psi)$. This action generalizes those used till now to construct extended
inflationary models\footnote{We point out that the more general action is
 $$
 {\cal A}=\int d^{4}x\sqrt{-g}\left[F(\varphi, \psi)R-V(\varphi, \psi)+A(\varphi, \psi)
 (\nabla\varphi)^2+B(\varphi, \psi)(\nabla\psi)^2\right]\,,
 $$
 but for the purpose of the paper, we will confine ourselves to the action (\ref{9})}.
 The Brans--Dicke action (\ref{1}) can be immediately recovered by using the
transformations (\ref{7}). In our units, the standard Newton coupling is recovered in
the limit $F(\varphi)+G(\psi)\to -1/2$. The field equations can be derived by varying
with respect to $g_{\mu\nu}$
\begin{equation}\label{10}
  [F(\varphi)+G(\psi)]\left(R_{\mu\nu}-\frac{1}{2}g_{\mu\nu}R\right)=
  T_{\mu\nu}^{(\varphi)}+T_{\mu\nu}^{(\psi)}\,.
\end{equation}
 In the right hand side of (\ref{10}) there is the effective stress--energy tensor containing
the nonminimal coupling contributions, the kinetic terms and the potentials of the
scalar fields $\varphi$ and $\psi$, that is
\begin{equation}\label{11}
 T_{\mu\nu}^{(\varphi)}=-\frac{1}{2}\varphi_{;\,\mu}\varphi_{;\,\nu}+
 \frac{1}{4}g_{\mu\nu}\varphi_{;\,\alpha}\varphi^{;\,\alpha}-
 \frac{1}{2}g_{\mu\nu}V(\varphi)-g_{\mu\nu}\Box F(\varphi)+F(\varphi)_{;\,\mu\nu}\,
\end{equation}
and analogously,
\begin{equation}\label{12}
 T_{\mu\nu}^{(\psi)}=-\frac{1}{2}\psi_{;\,\mu}\psi_{;\,\nu}+
 \frac{1}{4}g_{\mu\nu}\psi_{;\,\alpha}\psi^{;\,\alpha}-
 \frac{1}{2}g_{\mu\nu}W(\psi)-g_{\mu\nu}\Box G(\psi)+G(\psi)_{;\,\mu\nu}\,.
\end{equation}
 $\Box$ is the d'Alambertian operator. The variation with respect to $\varphi$ and
$\psi$ gives the Klein--Gordon equations
\begin{equation}\label{13}
  \Box \varphi-R\left(\frac{dF}{d\varphi}\right)+\frac{dV}{d\varphi}=0\,,
\end{equation}
and
\begin{equation}\label{14}
  \Box \psi-R\left(\frac{dG}{d\psi}\right)+\frac{dW}{d\psi}=0\,.
\end{equation}
 Their sum is equivalent to the contracted Bianchi identities \cite{cimento}. Let us now
take into account a FRW metric of the form
\begin{equation}\label{15}
  ds^2=dt^2-a^2\left[\frac{dr^2}{1-kr^2}+r^2d\Omega^2\right]
\end{equation}
and substitute it into the action (\ref{9}). Integrating by parts and eliminating the
boundary terms, we get the point-like Lagrangian
 \beq\label{16}
 {\cal L}=6\frac{dF}{d\varphi}a^{2}\dot{a}\dot{\varphi}+6Fa\dot{a}^{2}-6kFa+
 \frac{a^{3}\dot{\varphi}^{2}}{2}-a^{3}
 V(\varphi)+6\frac{dG}{d\psi}a^2\dot{a}\dot{\psi}+6Ga\dot{a}^2-6kGa+
 \frac{a^{3}\dot{\psi}^{2}}{2}-a^3W(\psi).
 \eeq
 The Euler--Lagrange equations, corresponding to the cosmological Einstein equations are
\begin{equation}\label{17}
  [F+G]\left[2\frac{\ddot{a}}{a}+\left(\frac{\dot{a}}{a}\right)^2+
  \frac{k}{a^2}\right]+2\left[\dot{\varphi}\frac{dF}{d\varphi}+\dot{\psi}\frac{dG}{d\psi}
  \right]\left(\frac{\dot{a}}{a}\right)+
\end{equation}
 $$
 +\left[\dot{\varphi}^2\frac{d^2F}{d\varphi^2}+\ddot{\varphi}\frac{dF}{d\varphi}+
 \dot{\psi}^2\frac{d^2F}{d\psi^2}+\ddot{\psi}\frac{dF}{d\psi}\right]-
 \frac{1}{2}\left[\frac{1}{2}(\dot{\varphi}^2+\dot{\psi}^2)-(V+W)\right]=0\,,
 $$
\begin{equation}\label{18}
 6[F+G]\left(\frac{\dot{a}}{a}\right)^2
 +6\left[\dot{\varphi}\frac{dF}{d\varphi}+
 \dot{\psi}\frac{dG}{d\psi}\right]\left(\frac{\dot{a}}{a}\right)
 +\frac{6k}{a^2}[F+G]+\frac{1}{2}(\dot{\varphi}^2+\dot{\psi}^2)+V+W=0\,,
\end{equation}
\begin{equation}\label{19}
  \ddot{\varphi}+3\left(\frac{\dot{a}}{a}\right)\dot{\varphi}+
  6\left[\frac{\ddot{a}}{a}+\left(\frac{\dot{a}}{a}\right)^2+
  \frac{k}{a^2}\right]\left(\frac{dF}{d\varphi}\right)+\frac{dV}{d\varphi}=0\,,
\end{equation}
\begin{equation}\label{20}
  \ddot{\psi}+3\left(\frac{\dot{a}}{a}\right)\dot{\psi}+
  6\left[\frac{\ddot{a}}{a}+\left(\frac{\dot{a}}{a}\right)^2+
  \frac{k}{a^2}\right]\left(\frac{dG}{d\psi}\right)+\frac{dW}{d\psi}=0\,,
\end{equation}
where Eq. (\ref{18}) is the energy constraint corresponding to the $(0, 0)$--Einstein
equation. Let us now go to solve the system (\ref{17})--(\ref{20}) by using the Noether
Symmetry Approach. The solutions strictly depend on the form of the functions $F, G, V$
and $W$. By the Noether symmetries it is possible to select these functions so that the
system (\ref{18})--(\ref{20}) can be reduced and then integrated.

\section{\normalsize\bf Selecting Couplings and Potentials by the Noether Symmetries}
\setcounter{equation}{0}

Given an undefined extended gravity theory, the existence of a Noether symmetry can
select the form of the coupling and the scalar field potential \cite{cimento} or the
form of the higher--order Lagrangian density, e.g. $f(R, \Box R)$ \cite{higher}.

At the same time, as we will show in the next section, the symmetry allows to reduce the
dynamical system by a cyclic variable making it easier to solve. Taking into account the
Lagrangian (\ref{16}), its configuration space is three--dimensional, $Q=\{a, \varphi,
\psi\}$. In the language of quantum cosmology, it can be identified with a {\it
minisuperspace} \cite{cimento}. The tangent space on which the Lagrangian (\ref{16}) is
defined is $TQ=\{a, \dot{a}, \varphi, \dot{\varphi}, \psi, \dot{\psi}\}$ so that the
lift vector $X$, the infinitesimal generator of symmetry, is
\begin{equation}\label{21}
  X=\alpha\frac{\partial}{\partial a}+\beta\frac{\partial}{\partial \varphi}+
  \gamma\frac{\partial}{\partial \psi}
  +\frac{d\alpha}{dt}\frac{\partial}{\partial \dot{a}}+
  \frac{d\beta}{dt}\frac{\partial}{\partial \dot{\varphi}}+
 \frac{d\gamma}{dt}\frac{\partial}{\partial \dot{\psi}}\,,
\end{equation}
 where $\alpha, \beta, \gamma$ are functions of $a, \varphi, \psi$. A Noether symmetry
exists if the condition
\begin{equation}\label{22}
  L_X{\cal L}=0\,,
\end{equation}
 is realized. $L_X$ is the Lie derivative wit respect to $X$.
Properly speaking, Eq. (\ref{22}) corresponds to the contraction
of the vector $X$ with the Lagrangian (\ref{16}). The constant of
motion connected to the Noether symmetry is nothing else but
\begin{equation}\label{23}
  \Sigma_0=i_X\vartheta_{{\cal L}}
\end{equation}
 where
\begin{equation}\label{24}
 \vartheta_{{\cal L}}=\frac{\partial {\cal L}}{\partial\dot{a}}\,da+
 \frac{\partial {\cal L}}{\partial\dot{\varphi}}\,d\varphi+
 \frac{\partial {\cal L}}{\partial\dot{\psi}}\,d\psi
\end{equation}
 is the Cartan one--form given by a Lagrangian ${\cal L}$ and
$i_X$ is the contraction with respect to $X$. The relation between
Eqs. (\ref{22}) and (\ref{23}) can be easily seen if the vector is
generally expressed as
\begin{equation}\label{21a}
  X=\alpha^i\,\frac{\partial}{\partial q^i}+\frac{d\alpha^i}{dt}\,
  \frac{\partial}{\partial\dot{q}^i}\,.
\end{equation}
 Using the Euler--Lagrange equations, it can be shown that
\cite{cimento}
\begin{equation}\label{25}
  \frac{d}{dt}\left(\alpha^i\,\frac{\partial {\cal L}}{\partial\dot{q}^i}\right)
 =L_X {\cal L}\,.
\end{equation}
 If the Noether symmetry exists, Eq. (\ref{25}) gives (\ref{22}).
In the Hamiltonian formalism, by a Legendre transformation, we get
\begin{equation}\label{26}
  {\cal
  H}=\dot{a}\pi_a+\dot{\varphi}\pi_{\varphi}+\dot{\psi}\pi_{\psi}-{\cal
  L}\,,
\end{equation}
 where $\pi_{q}\equiv \partial {\cal L}/\partial\dot{q}$, $q=\{a,
\varphi, \psi\}$ are the conjugate momenta. The phase--space
vector for the symmetry is now
\begin{equation}\label{27}
  \Gamma=\dot{a}\frac{\partial}{\partial a}+
  \dot{\varphi}\frac{\partial}{\partial \varphi}+
  \dot{a\psi}\frac{\partial}{\partial \psi}+
  \ddot{a}\frac{\partial}{\partial\dot{a}}+
  \ddot{\varphi}\frac{\partial}{\partial\dot{\varphi}}+
  \ddot{\psi}\frac{\partial}{\partial\dot{\psi}}\,,
\end{equation}
 and a Noether symmetry exists if
\begin{equation}
  L_{\Gamma}{\cal H}=0\,.
\end{equation}
 The conserved quantity (\ref{23}) and the Hamiltonian (\ref{26})
gives the Poisson brackets
\begin{equation}\label{28}
  \{\Sigma_0, {\cal H}\}=0\,.
\end{equation}
 Our issue is now to determine the functions $F, G, V, W$ by this
Noether symmetry technique. We shall adopt the Lagrangian
formalism. The condition (\ref{22}) gives the system of partial
differential equations
 \beq\label{30}
 F\left(\alpha + 2a\frac{\partial\alpha}{\partial a}\right)+
 G\left(\alpha + 2a\frac{\partial\alpha}{\partial a}\right)+
 a\left(\frac{dF}{d\varphi}\right)
 \left(\beta + a\frac{\partial\beta}{\partial
 a}\right)+
 a\left(\frac{dG}{d\psi}\right)\left(\gamma + a\frac{\partial\gamma}{\partial
 a}\right)=0\,,
 \eeq
 \beq\label{31}
 3\alpha+12\left(\frac{dF}{d\varphi}\right)\frac{\partial\alpha}{\partial\varphi}+2a\frac{\partial\beta}
 {\partial\varphi} = 0 \,{,}
 \eeq
 \beq\label{32}
 3\alpha+12\left(\frac{dG}{d\psi}\right)\frac{\partial\alpha}{\partial\psi}+
 2a\frac{\partial\gamma}{\partial\psi}= 0\,{,}
 \eeq
\begin{equation}\label{33}
  a\beta\frac{d^2F}{d\varphi^2}+\left(2\alpha+a\frac{\partial\alpha}{\partial a}+
  a\frac{\partial\beta}{\partial\varphi}\right)\left(\frac{dF}{d\varphi}\right)+
  2\frac{\partial\alpha}{\partial\varphi}F+\frac{a^2}{6}\frac{\partial\beta}{\partial
  a}+a\frac{\partial\gamma}{\partial\varphi}\left(\frac{dG}{d\psi}\right)
  +2\frac{\partial\alpha}{\partial\varphi}G=0\,,
\end{equation}
\begin{equation}\label{34}
  a\gamma\frac{d^2G}{d\psi^2}+\left(2\alpha+a\frac{\partial\alpha}{\partial
  a}+a\frac{\partial\gamma}{\partial\psi}\right)\left(\frac{dG}{d\psi}\right)+
  2\frac{\partial\alpha}{\partial\psi}G+\frac{a^2}{6}\frac{\partial\gamma}{\partial
  a}+a\frac{\partial\beta}{\partial\psi}\left(\frac{dF}{d\varphi}\right)+
  2\frac{\partial\alpha}{\partial\psi}F=0\,,
\end{equation}
\begin{equation}\label{35}
 6\frac{\partial\alpha}{\partial\varphi}\left(\frac{dG}{d\psi}\right)+
 6\frac{\partial\alpha}{\partial\psi}\left(\frac{dF}{d\varphi}\right)+
 \frac{\partial\beta}{\partial\psi}a+
 \frac{\partial\gamma}{\partial\varphi}a=0\,,
\end{equation}
 \beq\label{36}
 6k\left[\alpha F + \beta\left(\frac{dF}{d\varphi}\right)a+\alpha G+
 \gamma \left(\frac{dG}{d\psi}\right)a\right]+
 a^2\left[3\alpha V+\beta a\left(\frac{dV}{d\varphi}\right)+3\alpha W+
 \gamma a \left(\frac{dW }{d\psi}\right)\right]=0\,,
 \eeq
 obtained by equating to zero the second degree coefficients in
$\dot{a}$, $\dot{\varphi}$, $\dot{\psi}$. The number of these
equations is $1+n(n+1)/2$, where $n$ is the dimension of the
configuration space $Q$. The system (\ref{30})--(\ref{36}) is the
straightforward generalization of the system (5.30)--(5.34) in
\cite{cimento} for the case of two scalar fields.

The integration of (\ref{30})--(\ref{36}) gives as a result the
functions $\alpha(a, \varphi, \psi)$, $\beta(a, \varphi, \psi)$,
$\gamma(a, \varphi, \psi)$, $F(\varphi)$, $V(\varphi)$, $G(\psi)$,
$W(\psi)$. Solutions are not unique and the various cases which we have found
are summarized in Table I.

In Table II, being $G(\psi)=0$, the cases where only one
nonminimal coupling is present are summarized. The value of the
spatial curvature constant $k$ is also given. The quantities
$F_0$, $G_0$, $F_0'$, $\Lambda_{1, 2}$ are constants. Using these
results, the dynamical system (\ref{17})--(\ref{20}) can be
reduced since, as we shall show below, a change of variables can
be found where a cyclic coordinate is present. This feature allows
to integrate more simply the dynamics.

\section{\normalsize\bf The Cosmological Solutions}
\setcounter{equation}{0}

The existence of a Noether symmetry gives, in any case, a cyclic
variable so that the transformation
\begin{equation}\label{37}
  {\cal L}(a, \dot{a}, \varphi, \dot{\varphi}, \psi, \dot{\psi})
  \longleftrightarrow
 {\cal L}(w, \dot{w}, u, \dot{u}, \dot{z})
\end{equation}
 is always possible. If more than one symmetry exists, more than
one cyclic variable can be present \cite{cimento}. In a geometric
language, it is always possible to choose a new set $q_i=q_i(Q_k)$
$i, k=1,2,3,$ adapted to the foliation given by $X$
\begin{equation}\label{38}
 i_XdQ_3=1\,,\quad i_XdQ_j=0\,, \quad j=1,2,
\end{equation}
 where $i_X$, as before, is the contraction given by $X$ and
$dQ_j=(\partial Q_j/\partial q_i)dq_i$. Explicitly, in our case, Eqs.
(\ref{38}) become
\begin{equation}\label{39}
  \alpha\frac{\partial w}{\partial a}+\beta\frac{\partial w}{\partial
  \varphi}+\gamma\frac{\partial w}{\partial\psi}=0\,,
\end{equation}
\begin{equation}\label{40}
  \alpha\frac{\partial u}{\partial a}+\beta\frac{\partial u}{\partial
  \varphi}+\gamma\frac{\partial u}{\partial\psi}=0\,,
\end{equation}
\begin{equation}\label{41}
  \alpha\frac{\partial z}{\partial a}+\beta\frac{\partial z}{\partial
  \varphi}+\gamma\frac{\partial z}{\partial\psi}=1\,,
\end{equation}
 where $z$ is the cyclic variable. However, the transformation
(\ref{39})--(\ref{41}) are specified as soon as the functions
$\alpha$, $\beta$, $\gamma$ are given. As an example, let us take
into account Case 1 in Table I. Cases 2 and 3 of Table
I and 1,2,3 of Table II can be deduced from it. The system
(\ref{39})--(\ref{41}) is solved by the choice of the new
variables
\begin{equation}\label{42}
  w=a^3\psi^2\,, \quad u=a^3\varphi^2\,, \quad z=\ln a\,.
\end{equation}
 For the scalar--field functions, as we said, we choose Case 1 in Table
I. Lagrangian (\ref{16}) becomes
\begin{equation}\label{43}
  {\cal L}=
  6F_0\dot{z}(\dot{w}-2\dot{z}w)+\frac{1}{8w}(\dot{w}-3\dot{z}w)^2-V_0w+
  6G_0\dot{z}(\dot{u}-2\dot{z}u)+\frac{1}{8w}(\dot{u}-3\dot{z}u)^2-W_0u
 \end{equation}
while the equations of motion are
\begin{equation}\label{44}
  6(8G_0-1)\ddot{z}+3(32G_0-3)\dot{z}^2+2\left(\frac{\ddot{u}}{u}-
  \frac{\dot{u}^2}{2u^2}\right)+8W_0=0\,,
\end{equation}
\begin{equation}\label{45}
  6(8F_0-1)\ddot{z}+3(32F_0-3)\dot{z}^2+2\left(\frac{\ddot{w}}{w}-
  \frac{\dot{w}^2}{2w^2}\right)+8V_0=0\,,
\end{equation}
\begin{equation}\label{46}
 (8F_0-1)\dot{w}-(32F_0-3)\dot{z}w+(8G_0-1)\dot{u}-(32G_0-3)\dot{z}u=\Sigma_0\,
\end{equation}
\begin{equation}\label{47}
   6(8F_0-1)\dot{z}\dot{w}-3(32F_0-3)
  \dot{z}^2w+\frac{\dot{w}^2}{w}+8V_0w+
  6(8G_0-1)\dot{z}\dot{u}-3(32G_0-3)\dot{z}^2u+\frac{\dot{u}^2}{u}+8W_0u=0\,,
\end{equation}
 where, clearly, $z$ is the cyclic variable and $\Sigma_0$ is the
constant of motion connected to $z$. With respect to the system
(\ref{17})--(\ref{20}), system (\ref{44})--(\ref{47}) is reduced
and it is highly symmetric due to the functions $F(\varphi), G(\psi),
V(\varphi), W(\psi)$
selected by the Noether symmetry.

Since the role of the two fields is completely symmetric, we can
suppose that, depending on the value of the parameters, one can select two
regimes of physical interest. For example, in the first case,
dynamics is $\varphi$ ($u$)--dominated, in the second case, it is
$\psi$ ($w$)--dominated. The relation among the initial data is
given by the energy condition (\ref{47}). We get in the $\varphi$
($u$)--dominated regime
\begin{equation}\label{48}
 u(t)=c_1\exp[\lambda_{1}t]+c_{2}\exp[-\lambda_{1}t]\,,
\end{equation}
\begin{equation}\label{49}
 z(t)=z_1\arctan\left(\sqrt{\frac{c_{1}}{c_{2}}}\exp[\lambda_{1}t]\right)
 +z_{2}\ln|c_1\exp[\lambda_{1}t]+c_{2}\exp[-\lambda_{1}t]|+z_{0}\,,
\end{equation}
for $c_{1}c_{2}>0$, and
\begin{equation}\label{50}
 z(t)=-z_1\mbox{arctanh}\left(\sqrt{\vert\frac{c_{1}}{c_{2}}\vert}
 \exp[-\lambda_{1}t]\right)
 +z_{2}\ln|c_1\exp[\lambda_{1}t]+c_{2}\exp[-\lambda_{1}t]|+z_{0}\,,
\end{equation}
for $c_{1}c_{2}<0$, while in the $\psi$ ($w$)--dominated regime
\begin{equation}\label{51}
 w(t)=c_3\exp[\lambda_{2}t]+c_{4}\exp[-\lambda_{2}t]\,,
\end{equation}
\begin{equation}\label{52}
 z(t)=z_3\arctan\left(\sqrt{\frac{c_{3}}{c_{4}}}\exp[\lambda_{2}t]\right)
 +z_{4}\ln|c_3\exp[\lambda_{2}t]+c_{4}\exp[-\lambda_{2}t]|+z_{0}\,,
\end{equation}
for $c_{3}c_{4}>0$, and
\begin{equation}\label{53}
 z(t)=-z_3\mbox{arctanh}\left(\sqrt{\vert\frac{c_{3}}{c_{4}}\vert}
 \exp[-\lambda_{2}t]\right)
 +z_{4}\ln|c_3\exp[\lambda_{2}t]+c_{4}\exp[-\lambda_{2}t]|+z_{0}\,,
\end{equation}
for $c_{3}c_{4}<0$. The constants $z_{0}$,
$c_{i}, z_{j}$, $i, j=1, \ldots ,4$
are integration constants while
\begin{equation}\label{54}
\lambda_{1}=\frac{V_{0}(32F_{0}-3)}{4F_{0}(12F_{0}-1)}\,,\quad
\lambda_{2}=\frac{W_{0}(32G_{0}-3)}{4G_{0}(12G_{0}-1)}\,,
\end{equation}
assume the role of cosmological constants depending on the
couplings and the potentials.

Inverting the relations (\ref{42}), we get
\begin{equation}\label{55}
  a(t)=a_0\left(c_1\exp[\lambda_1t]+c_2\exp[-\lambda_1t]\right)\exp\left[
  z_1\arctan\left(\sqrt{\frac{c_1}{c_2}}\exp[\lambda_1t]\right)\right]\,,
\end{equation}
\begin{equation}\label{56}
  \varphi(t)=\varphi_0\left(c_1\exp[\lambda_1t]+c_2\exp[-\lambda_1t]
  \right)^{-2}
  \left(\exp\left[
  z_1\arctan\left(\sqrt{\frac{c_1}{c_2}}\exp[\lambda_1t]\right)\right]\right)^{-3}\,,
\end{equation}
for $c_1 c_2>0$, and
\begin{equation}\label{57}
  a(t)=a_0\left(c_1\exp[\lambda_1t]+c_2\exp[-\lambda_2t]\right)\exp\left[
  z_1\mbox{arctanh}\left(\sqrt{\vert\frac{c_1}{c_2}\vert}\exp[\lambda_1t]\right)\right]\,,
\end{equation}
\begin{equation}\label{58}
  \varphi(t)=\varphi_0\left(c_1\exp[\lambda_1t]+c_2\exp[-\lambda_2t]\right)^{-2}
  \left(\exp\left[
  z_1\mbox{arctanh}\left(\sqrt{\frac{c_1}{c_2}}\exp[\lambda_1t]\right)\right]\right)^{-3}\,,
\end{equation}
for $c_1c_2<0$ in the $\varphi$--dominated regime.

The situation is analogous in the $\psi$--dominate regime, but the
constants $\lambda_1$, $z_1$, $c_{1, 2}$ and $\varphi_0$ have to
be substituted with $\lambda_2$, $z_2$, $c_{3, 4}$ and $\psi_0$.

It is easy to see hat we have two inflationary eras. Their
durations are ruled by the parameter $\lambda_1, \lambda_2$ which
strictly depends on the strength of the couplings. Another
interesting particular solution is
\begin{equation}\label{59}
  a(t)=a_0e^{z_0t}\,,\quad
  \varphi(t)=\varphi_0\exp
  \left\{\frac{4F_0z_0}{8F_0-1}t\right\}\,,\quad
  \psi(t)=\psi_0\exp
  \left\{\frac{4G_0z_0}{8G_0-1}t\right\}\,,\quad
\end{equation}
where
\begin{equation}\label{60}
  V_0=-\frac{z_0^2}{8(8F_0-1)}F_0
  \left(F_0-\frac{1}{12}\right)
  \left(F_0-\frac{1}{10.6}\right)
\end{equation}
\begin{equation}\label{61}
  W_0=-\frac{z_0^2}{8(8G_0-1)}G_0
  \left(G_0-\frac{1}{12}\right)
  \left(G_0-\frac{1}{10.6}\right)\,.
\end{equation}
Also here the inflationary behaviour is clear.

By using similar arguments, we can analyze  Case 2 in Tab.I
 Here the potential terms cancel each other in the Lagrangian
(\ref{43}). We get the power--law solution
\begin{equation}\label{62}
  a(t)=a_0t^n\,,\quad \varphi(t)=\varphi_0 t^{-n}\,{,} \quad
 \psi(t)=\psi_0t^{-n}\,{,}
\end{equation}
which is particularly useful for extended inflation being $n$ an
arbitrary constant depending on the initial conditions. A similar
situation holds in Case 4, which is a minimally coupled case
with two fields and two cosmological constants. For $k=0$ and
$\Lambda_1=-|\Lambda|_2$, one finds
 \beq\label{63}
 a(t)=a_0t^n\,{,}\quad \varphi(t)=\frac{\Sigma_0}{a_0^3(1-3n)}\,
 t^{1-3n}+\varphi_0\,{,}\quad \psi(t)=\frac{\Sigma_1}{a_0^3(1-3n)}\,
 t^{1-3n}+\psi_1\,{,}
 \eeq
 for $n\neq 1/3$, and
 \beq\label{64}
 a(t)=a_0t^{1/3}\,{,}\quad \varphi(t)=\frac{\Sigma_0}{a_0^3}\ln
 t+\varphi_0\,{,}\quad \psi(t)=\frac{\Sigma_1}{a_0^3}\ln
 t+\psi_0\,{,}
 \eeq
 when $n=1/3$. In this case, two Noether symmetries are present and
they assign the value of gravitational constant being
 \beq
 F_0+G_0=-\frac{3(\Sigma_0^2+\Sigma_1^2)}{4a_0^6}\,{.}
 \eeq
 Let us now analyse, in detail, Case 5. Without loosing of generality,
we can assume $F_0'=F_0=0$, $\gamma_{0}=1$  and studying the
couplings $F(\varphi)=\varphi^2/12$ and $G(\psi)=G_0$. Lagrangian
(\ref{16}) becomes
\begin{equation}\label{65}
  {\cal L}=\left(\frac{\varphi^2}{2}+6G_0\right)a\dot{a}^2+
  a^2\dot{a}\varphi\dot{\varphi}+a^3\left(\frac{\dot{\varphi}^2}{2}-\Lambda\right)
  +\frac{a^3\dot{\psi}}{2}\,
\end{equation}
where $\Lambda=\Lambda_1+\Lambda_2$. Clearly $\psi$ is the cyclic
variable. Eqs. (\ref{39})--(\ref{41}) are satisfied by
\begin{equation}\label{66}
  w=a\,,\quad u=a\varphi-\psi\,, \quad z=a\varphi\,.
\end{equation}
 With the further change
\begin{equation}\label{67}
  \chi=z-u\,,
\end{equation}
Lagrangian (\ref{65}) reads
\begin{equation}\label{68}
  {\cal L}= \left(6G_0\dot{w}^2+\frac{\dot{z}^2}{2}\right)w+
  \left(\frac{\dot{\chi}^2}{2}-\Lambda\right)w^3\,,
\end{equation}
where two cyclic variables appear. The dynamical system is
\begin{equation}\label{69}
  6G_0(2\ddot{w}w+\dot{w}^2)=\frac{\dot{z}^2}{2}+3w^2
  \left(\frac{\dot{\chi}^2}{2}-\Lambda\right)\,,
\end{equation}
\begin{equation}\label{70}
  \dot{\chi}w^3=\Sigma_1\,,
\end{equation}
\begin{equation}\label{71}
  \dot{z}w=\Sigma_2\,,
\end{equation}
\begin{equation}\label{72}
  \dot{z}^2+12G_0\dot{w}^2+\dot{\chi}^2w^2+2\Lambda w^2=0\,,
\end{equation}
whose general solution is given by the elliptic integral
\begin{equation}\label{73}
  \int \frac{w^2dw}{\sqrt{A_1w^6+A_2w^2+A_3}}=\pm t\,,
\end{equation}
where
\begin{equation}\label{74}
  A_1=-\frac{A_1}{6G_0}\,,\quad A_2=-\frac{\Sigma_2^2}{6G_0}\,,
  \quad A_3=-\frac{\Sigma_1^2}{12G_0}\,.
\end{equation}
In the particular case where $A_2=0$, we get the explicit solution
\begin{equation}\label{75}
  a(t)=a_0\sqrt[3]{\frac{\Sigma_1^2}{2\Lambda}\sinh (\pm
  3\sqrt{A_1}\,t)}\,,
\end{equation}
\begin{equation}\label{76}
  \varphi(t)=\frac{\varphi_0}{\sqrt[3]{\frac{\Sigma_1^2}{2\Lambda}\sinh (\pm
  3\sqrt{A_1}\,t)}}\,,
\end{equation}
\begin{equation}\label{77}
  \psi(t)=-\frac{\cosh (\pm 3\sqrt{A_1})\, t}{3\sqrt{A_1}\,\sinh^2(\pm \sqrt{A_1}\,t)}
  -\frac{1}{6\sqrt{A_1}}\,\ln\tanh\left(\frac{\pm 3\sqrt{A_1}\,
  t}{2}\right)+\psi_0\,,
\end{equation}
which asymptotically gives a de Sitter behaviour. A last
interesting case is 8 in Tab.I, which can be assigned by the
functions
 $$
  F=F_0\,, \quad V(\varphi)=\Lambda\,, \quad k=-1\,,
  $$
\begin{equation}\label{78}
 G(\psi)=G_0+\frac{\psi^2}{16}+
 \left(\frac{\varphi_0^2}{12\psi_0^4}\right)\psi^4\,,
 \quad W(\psi)=\frac{\psi^4}{4\psi_0^2}-\Lambda\,,
\end{equation}
being $G(\psi)$ and $W(\psi)$ free for the Noether symmetry. The
model is relevant for hyper-extended inflation (see e.g.
\cite{accetta}). Power law solutions like in \cite{accetta,liddle}
are easily found.

The cases in Tab.II are essentially subcases of those discussed
above.

\section{\normalsize\bf Inflation and  graceful exit}
\setcounter{equation}{0}

As we discussed in Introduction, the goal to get a sufficient inflationary period and
then to exit from it, without imposing any particular fine tuning, can be achieved by
assuming the variation of the bubble nucleation rate $\epsilon$. However, we are taking
into account a first order phase transition after which we recover a Friedman stage
\cite{kolb}. In principle, being $\epsilon=\lambda/H^4$, we can expect the variation of
both $\lambda$ and $H$. The form of $\lambda$ strictly depends on the form of the theory
and, as it is discussed in \cite{vadas}, it is time independent toward the late times,
if we are dealing with a Brans--Dicke theory. In our cases, by using (\ref{7}), we get
that most of the couplings selected by the existence of Noether symmetry can be recast
in a Brans--Dicke--like form.

In spite of the variation of the effective gravitational coupling, it is reasonable to
assume $\lambda$ to be approximatively constant \cite{kolb}, so that the mechanism of
the graceful exit can be essentially connected to the variation of $H$.

However, this argument does not work for more general classes of theories, as
hyperextended inflation \cite{accetta}, where $\omega(\phi)$ is not a constant. Among
the cases in Tabs. I and II, we have also couplings of the form
\begin{equation}
\omega(\phi)=\frac{F(\varphi)}{2(dF/d\varphi)^2}=\frac{\frac{1}{12}\varphi^2+F_{0}'\varphi+
F_{0}}{2\left[\frac{1}{6}\varphi+F_{0}'\right]^2}\,,
\end{equation}
see e.g. the cases 5 in Tab.I and 4 in Tab.II. This situation deserves more attention
since we can distinguish a regime where we match a sort of hyperextended inflation
$(\varphi\rightarrow 0)$ and a regime where the extended inflationary scheme is
recovered $(\varphi\rightarrow\infty)$. In any case, the microwave background bounds
have to be satisfied, as discussed in \cite {liddle}. Furthermore, taking into account
double--field models, the contributions to $\lambda$ come from $\varphi$ and $\psi$.
Being both fields nonminimally coupled, and from the forms of couplings selected by the
Noether symmetry, we are dealing with a double Brans--Dicke--like theory where the
extended inflationary mechanism is improved. Looking at the solutions of previous
section, we can have double inflationary stages ruled by the parameters of couplings and
self--interaction potentials (see e.g. (\ref{55})--(\ref{58}) or (\ref{62})). This
situation is extremely interesting since ''very" large scale structure and large scale
structure can be selected by these inflationary phases. In fact, we can have ''two"
first--order phase transitions and then ''two" bubble nucleations where the size of
bubbles is given by the coupling parameters. In other words, we can expect two graceful
exits given by the superposition of two extended inflationary phases. To be more
precise, at a given time $t>t_{0}$ after nucleation, the ''comoving" bubble radius is
\begin{equation}
r(t,t_0)=\int^{t}_{t_{0}} dt a(t)^{-1}\,,
\end{equation}
while the ''physical" size of the bubble is
\begin{equation}
{\cal R}(t,t_0)=a(t)r(t,t_0)\,.
\end{equation}
When $t\rightarrow\infty$, the form of $a(t)$ selects the size of the bubble. In the
cases (\ref{55}),(\ref{58}) and (\ref{59}), this size is finite since the asymptotic
behaviour is $a(t)\sim\exp H_0 t$. For power law behaviours, the growth of the bubble
size is linear.

Besides, we have a variation of the Hubble parameter in most of the cases we dealt with:
in (\ref{55}), (\ref{57}), and (\ref{75}), it converges to a constant for
$t\rightarrow\infty$, in (\ref{59}), it is exactly a constant, in the other cases, it is
$H\sim t^{-1}$.

Graceful exit is achieved if, being $\lambda$ a constant, $\epsilon$ is less than
$\epsilon_{crit}$ during inflation and, after bubble nucleation, $\epsilon
>\epsilon_{crit}$.  In our cases, $H$ is the key parameter  which governs
the behaviour of $\epsilon$. For power--law solutions, as for standard extended
inflationary models, the graceful exit is easily recovered (see Eqs(\ref{62}) and
(\ref{63})). In the asymptotically exponential cases, the parameter $\epsilon$ goes to a
constant for $t\rightarrow\infty$ and the graceful exit problem has no solution. In
fact, the function $H$, calculated from (\ref{55}) and (\ref{57}), is a sort of step
function with two different constant values at $t\rightarrow\pm\infty$. The related
$H^{-4}$ has a singularity in the origin which does not allow a graceful exit from
inflation. The situation for the solution (\ref{75}) is similar.

\section{\normalsize\bf  Conclusions}
\setcounter{equation}{0}

In this paper, we derived exact cosmological solutions in double
scalar--tensor gravity theories by the general approach of
searching for Noether symmetries. This work generalizes those in
\cite{modak,capozziello,cimento}. The couplings and the
potentials of both scalar fields are connected with the existence
of the symmetries, and the solutions of dynamics furnish
power law or de Sitter
evolutions. As a consequence, in all the above cases it is easy
to calculate the bubble nucleation rate $\epsilon=\lambda/H^4$
to test if one succeeds in graceful exit. Depending on the value
of intervening parameters, this can be accomplished in several cases.

Furthermore, being in principle, both fields nonminimally coupled
and self--interacting with a potential, their role is mixed and it
is not possible to distinguish, a priori, a Brans--Dicke field and
an inflaton field as in other extended inflationary models. This
distinction seems, in our opinion, rather artificial in view of a
stochastic approach to the fundamental interactions where the
effective role of the various fields is distinguishable only in
the low energy limit (see \cite{ottewill} and reference therein)
and there is no reason why a field should interact with the
gravitational field and the other one not.

Another remark deserves double inflation which is ruled by the
parameters of the theory and, then, by the Noether symmetry. As it
is well known this feature is of extreme interest in perturbation
theory since it can furnish the seeds for the formation of
structures at large and at very large scales.
As we have seen, it is very common in our approach and it could
contribute to the enhancement of a successful extended inflation.

Finally, we want to stress the fact that the standard Newtonian coupling
can be recovered in several of the above models, being
\begin{equation}
  G_{eff}=-\frac{1}{2[F(\varphi)+G(\psi)]}\,,
\end{equation}
so that as soon as $F(\varphi)\to F_0$ and/or $G(\psi)\to G_0$,
general relativity is restored (in our units $F_0+G_0\rightarrow -1/2$)
and both fields can contribute to
its recovering. This means that in an accurate setting of the models,
one could succeed both in graceful exit and in recovering of
standard gravity.

\vspace{4. mm}

\begin{center}
Table I -- {\it Symmetries in double nonminimally coupled models.}
\end{center}
\begin{center}
\begin{tabular}{ccccccccc} \hline\hline
N. & $\alpha$ & $\beta$ & $\gamma$ & $F(\varphi)$ & $G(\psi)$ & $V(\varphi)$ & $W(\psi)$
& $k$  \\ \hline
 1 & $a$ & $-3\varphi/2$ & $-3\psi/2$ & $F_0\varphi^2$ & $G_0\psi^2$ &
 $V_0\varphi^2$ & $W_0\psi^2$ & 0 \\ \hline
 2 & $a$ & $-3\varphi/2$ & $-3\psi/2$ & $F_0\varphi^2$ & $G_0\psi^2$ &
 $\Lambda$ & $-\Lambda$ & 0 \\ \hline
 3 & $a$ & $-3\varphi/2$ & $-3\psi/2$ & $F_0\varphi^2$ & $G_0\psi^2$ &
 0 & $W_0\psi^2$ & 0 \\ \hline
 4 & 0 & 0 & $\gamma_0$ & $F_0$ & $G_0$ & $\Lambda_1$ &
 $\Lambda_2$ & $\forall k$ \\ \hline
 5 & 0 & $1/a$ & $\gamma_0$ & $
 \frac{1}{12}\varphi^2+F_0'\varphi+F_0$ & $G_0$ & $\Lambda_1$ &
 $\Lambda_2$ & 0 \\ \hline
 6 & 0 & $1/a$ & $\gamma_0$ & $
 \frac{1}{12}\varphi^2+F_0'\varphi+F_0$ & $G_0$ & $0$ &
 $\Lambda$ & 0 \\ \hline
 7 & 0 & $1/a$ & $\gamma_0$ & $
 \frac{1}{12}\varphi^2$ & $G_0$ & $\Lambda_1$ &
 $\Lambda_2$ & 0 \\ \hline
 8 & 0 & $\beta_0$ & 0 & $F_0$ & $G(\psi)$ & $ 0, \Lambda$ &
 $W(\psi)$ & $\forall k$ \\ \hline
 9 & 0 & $1/a$ & 0 & $\frac{1}{12}\varphi^2+F_0'\varphi+F_0$ &
 $G(\psi)$ & $0, \Lambda$ & $W(\psi)$ & 0 \\ \hline\hline
 \end{tabular}
 \end{center}

\vspace{1. cm}

\begin{center}
Table II -- {\it Symmetries in single nonminimally coupled models.}
\end{center}
\begin{center}
\begin{tabular}{cccccccc}\hline\hline
N.  & $\alpha$  &  $\beta$  &  $\gamma$  & $F(\varphi)$ &
$V(\varphi)$  &  $W(\psi)$ & $k$ \\ \hline\hline
 1 & $a$  &  $-3/2\varphi$  & $-3\psi/2$ & $F_0\varphi^2$ &
 $V_0\varphi^2$ & $W_0\psi^2$ & 0 \\ \hline
 2 & $a$  &  $-3\varphi/2$  & $-3\psi/2$ & $F_0\varphi^2$ &
 $0$ & $W_0\psi^2$ & 0 \\ \hline
 3 & $a$  &  $-3\varphi/2$  & $-3\psi/2$ & $F_0\varphi^2$ &
 $\Lambda$ & $-\Lambda$ & 0 \\ \hline
 4 & 0 & $1/a$ & $\gamma_0$ &
 $\frac{1}{12}\varphi^2+F_0'\varphi+F_0$ & 0 & $\Lambda$ & 0 \\
                                                    \hline
 5 &  0 & $1/a$ & $\gamma_0$ &
 $\frac{1}{12}\varphi^2+F_0'\varphi+F_0$ & $\Lambda_1$ & $\Lambda_2$ & 0 \\
                                                    \hline
 6 & 0 & 0 & $\gamma_0$ & $F_0$ & $\Lambda_1$ & $ \Lambda_2$ &
 $\forall k$ \\ \hline\hline
 \end{tabular}
 \end{center}

\bigskip

\centerline{\bf Acknowledgments}

The authors thank the referee for the useful comments. Research supported by MURST fund
40\% and 60\% art. 65 D.P.R. 382/80. GL acknowledges UE (P.O.M. 1994/1999) for financial
support.

\end{document}